\documentclass[useAMS,usenatbib]{mn2e}
\usepackage{graphicx}
\usepackage{natbib}

\title[Hard gamma-ray spectra of blazars due to 
internal absorption]{Formation of hard VHE gamma-ray spectra of blazars due to 
internal photon-photon absorption}

\author[Aharonian, Khangulyan \& Costamante]{Felix A.  Aharonian\thanks{E-mail:felix.aharonian@dias.ie}${}^{1,2}$, D.Khangulyan${}^2$ \& L. Costamante${}^3$\\
${}^1$Dublin Institute for Advanced Studies, 31 Fitzwilliam Place,
Dublin 2, Ireland\\
${}^2$Max Planck Institut f\"ur Kernphysik, 
Saupfercheckweg 1, D69117 Heidelberg, Germany\\
${}^3$Stanford University, W.W. Hansen Experimental Physics Laboratory \&\\
Kavli Institute for Particle Astrophysics and Cosmology,
Stanford, CA 94305-4085, USA}

\begin{document}

\date{Accepted. Received; in original form}

\pagerange{\pageref{firstpage}--\pageref{lastpage}} \pubyear{}

\maketitle

\label{firstpage}

\begin{abstract}
The energy spectra of TeV gamma-rays from blazars,
after being corrected for intergalatic absorption 
in the Extragalactic Background Light (EBL), appear unusually
hard, a fact  that poses challenges 
to the conventional models of particle acceleration in TeV 
blazars and/or to the EBL models. In this paper we show that the 
internal absorption of gamma-rays caused by interactions with dense 
narrow-band radiation fields in the vicinity of 
compact gamma-ray production 
regions can lead to the formation of gamma-ray spectra of an almost 
arbitrary hardness. This allows significant relaxation of  
the current tight constraints on particle acceleration and radiation 
models,  although at the expense of enhanced requirements 
to the available nonthermal energy budget. 
The latter, however, is not a critical issue, as long as it 
can be largely compensated 
by the Doppler boosting, assuming very large ($\geq 30$) Doppler 
factors of the relativistically moving gamma-ray production regions.
The suggested scenario of formation of hard gamma-ray spectra  
predicts detectable synchrotron radiation of secondary  
electron-positron pairs which might require 
a revision of the  current ``standard paradigm''  
of  spectral energy distributions of gamma-ray blazars. If the primary 
gamma-rays are of hadronic origin related to $pp$ or $p \gamma$ interactions, the ``internal gamma-ray absorption'' model 
predicts neutrino fluxes close to the detection threshold of the 
next  generation high energy neutrino detectors. 
\end{abstract}

keywords: {BL Lacertae objects: general, gamma-rays:
theory, gamma-rays:
observations,
diffuse radiation}

\section{Introduction}

The recent reports on detections of very high energy (VHE)  
gamma-rays from blazars with redshifts $z \geq 0.1$ (for a review see e.g.
Hinton (2007)) initiated renewed debates on the interpretation
of TeV gamma-ray spectra of blazars, in particular in the context of the 
level of the diffuse extragalactic background radiation 
at optical and infrared wavelengths, often called also 
as Extragalactic Background Light (EBL). Initially, the 
tight link  between these two topics - TeV blazars and EBL - 
became a subject of hot discussions  prompted 
by multi (up to 20) TeV gamma-rays detected from a 
nearby BL Lac object,  Mkn~501  \cite{mkn501_HEGRA},  and by
the reports claiming detection of high  fluxes of EBL 
at far infrared  wavelengths \cite{Hauser,Schlegel,Lagage,Fink}. 
However, it was quickly recognised that these two claims 
hardly could be  compatible within any standard  model of TeV blazars (see, for a review, Aharonian (2001)). 

A distinct feature of extragalactic gamma-ray 
astronomy is that VHE  gamma-rays emitted by distant 
($\geq 100 \ \rm Mpc$) objects arrive after significant absorption 
caused by their interactions with EBL via 
the process $\gamma \gamma \rightarrow e^+ e^-$ \cite{Nikishov,Jelley,Gould}. 
The reconstructed, i.e. the absorption-corrected 
gamma-ray spectrum from a source at a redshift z, 
$J_0(E)=J_{\rm obs}(E) e^{\tau(E,z)}$ 
depends on the flux and energy spectrum of EBL through the optical 
depth $\tau(E,z)$. Thus, at energies where $\tau(E,z) \geq 1$, the 
primary gamma-rays suffer strong spectral deformation.

The EBL consists of two emission components 
produced by stars and partly absorbed/re-emitted by dust throughout the 
entire history of galaxy evolution. 
As a result, two distinct bumps are expected in the 
spectral energy distribution (SED) of EBL 
at near infrared (NIR) and far infrared (FIR) 
wavelengths, with a mid-infrared (MIR) 
``valley'' between these two bumps 
(see e.g. Hauser and Dwek (2001)). 
Generally, for almost all EBL models,  $\tau(E)$ is a strong function 
of energy  below 1 TeV and above 10 TeV; between 1 and 10 TeV the 
energy-dependence of  $\tau(E)$ is much weaker \cite{Hamburg01}. 
Consequently, one  should expect 
significant distortion of the VHE spectra of blazars at energies 
below 1 TeV and  above 10 TeV, provided that at these 
energies $\tau \geq 1$. 
One can re-formulate  this statement in a different way. 
Namely, for a standard (``decent'') intrinsic gamma-ray spectrum, 
the observer should detect  very soft (steep) spectra 
at energies below 1 TeV and above 10 TeV from objects for which 
$\tau \geq 1$  at corresponding energies. This condition 
is safely satisfied, given the constraints on the minimum  EBL flux
imposed by  galaxy counts, 
for blazars with redshifts $z \geq 0.15$ like 1ES~1101-232 and 
for nearby objects with $z \sim 0.03$ like Mkn~501. 
Even though the \textit{detected} gamma-ray spectra from both objects 
in the corresponding energy intervals are indeed quite steep   
with a photon index $\sim 3$ \cite{mkn501_HEGRA,1101_HESS}, 
they appear not sufficiently steep to compensate the function 
$f(E)=e^{\tau(E)}$, and thus to prevent a robust 
conclusion that the \textit{intrinsic} VHE gamma-ray spectra 
of these blazars are unusually hard. 

In the case of Mkn 501, the intrinsic spectrum has  
a ``non-standard'' shape with a possible pile-up above 
10 TeV which has been interpreted as a ``IR background - TeV gamma-ray 
crisis'' \cite{IRcrisis}
or a need to invoke dramatic assumptions like a violation of 
the Lorentz invariance (see e.g. Kifune (1999)). However, a more 
pragmatic view 
which presently dominates in both  infrared and gamma-ray
astronomical communities, treats this ``crisis''  
as somewhat exaggerated, especially given the ambiguity of 
extraction of the  truly diffuse  extragalactic FIR component from the 
much higher backgrounds of local origin (see e.g. Hauser and Dwek (2001)). 
Nevertheless, the recently reported low limits on the EBL at 
mid infrared wavelengths from the Spitzer deep cosmological 
surveys appeared quite high, for example 
at 70 $\mu \rm m$  the EBL flux should exceed  
$\geq 7.1 \pm 1.0 \ \rm nW/m^2 s$ \cite{Dole}. This 
implies that the problem is not yet over, and 
one may still face a challenge with the 
interpretation of the energy spectra of Mkn 501 
and Mkn 421 in the multi-TeV energy domain.

On the other hand, the recent detections of TeV
gamma-rays from blazars with redshifts 
$z \geq  0.15$  renewed the potential problems and 
challenges for standard models of TeV blazars. This time 
the issue has a more solid experimental background, because 
the gamma-ray spectra corrected for the intergalactic absorption appear very
hard (``harder than should be'') even for the minimum possible EBL fluxes 
at optical and NIR  wavelengths. 
Namely, the HESS collaboration reported, based on the detection of TeV gamma-rays from 
the BL Lac object 1ES~1101-232, that any significant deviation from the 
lower limits of  EBL determined by the integrated light of galaxies 
resolved by the Hubble telescope \cite{Madau}, 
would lead  to very hard intrinsic gamma-ray spectrum with a slope characterized by a 
photon index $\Gamma_0 \leq 1.5$ \cite{1101_HESS}. The analysis based on a larger sample 
of TeV blazars leads to the same conclusion \cite{Mazin}.
Recently, the HESS collaboration reported detection of multi-TeV gamma-rays 
from 1ES 0229+200, a BL Lac object located at a redshift z=0.1396 \cite{0229_HESS}. 
It is remarkable that the \textit{detected} 
hard gamma-ray spectrum of this source 
with a photon index  $\Gamma_{\rm obs} \sim 2.5$ extends up to 15 TeV. 
This, to a certain extent surprising result 
can be explained by the shape of the energy flux of EBL which  
between the NIR and MIR  bands  
is expected to be  proportional to $\lambda^{-1}$ \cite{Hamburg01}. 
Yet, the absolute EBL flux, derived from a rather conservative 
assumption that  the photon index of the intrinsic spectrum of TeV 
gamma-rays  does not  exceed $1.5$, 
appears again close to the EBL lower limit,
this time at MIR ($\approx 2-3 \ \rm nW/m^2 sr$ at $10 \ \rm \mu m$), derived  
from the Spitzer galaxy counts \cite{Fazio,Dole}.
Thus, the gamma-ray observations of 1ES~1101-232 and 
1ES 0229+200 can be interpreted  as an argument that the galaxies
resolved by the Hubble and Spitzer telescopes provide the bulk of the 
EBL flux  from optical to mid infrared wavelengths. 
Given the importance of such a  
statement, in particular for understanding of 
contribution of the first stars to the EBL 
(see e.g. Kashlinsky (2005); Mapelli et al. (2006)), 
it is essential to explore alternative 
ways of explanation of  very hard  intrinsic gamma-ray spectra or even 
sharper spectral features (like pile-ups) in TeV blazars. 
In this context, recently  some extreme assumptions regarding 
the distributions of accelerated particles 
have been proposed. In particular,
Katarzynski et al  (2006) argued that a gamma-ray spectrum as hard 
as $\Gamma_0 \sim 0.7$ can be formed in a SSC model assuming 
a narrow parent electron distribution, e.g. power-law  
within $E_1$ and $E_2$, with  a low-energy cutoff 
$E_1$ not much smaller than the high energy cutoff, $E_2$.  
In similar lines, Stecker et al (2007) argued 
that electron spectra with power law index $\leq 1$ 
can be accommodated within the  models of 
relativistic shock acceleration . 
It should be noted, however, 
that in compact objects relativistic electrons 
usually suffer very fast synchrotron losses, 
therefore the assumptions about hard 
electron \textit{acceleration } cannot 
yet guarantee hard gamma-ray spectra. 
Indeed, the radiatively cooled electron spectrum 
cannot be harder than $dN/dE \propto E^{-2}$, independent 
of the initial (acceleration) spectrum (see e.g. Aharonian 2004). 
If so, the inverse Compton  scattering  would result in 
a gamma-ray spectrum steeper than $E^{-1.5}$. In fact, the 
Klein-Nishina effect makes the spectrum even steeper. 
In principle, one can avoid the synchrotron  
cooling of electrons, e.g. in a cold ultrarelativistic 
wind. However such a hypothesis suggested for 
Mkn~501\cite{ColdWind}, in analogy with pulsar winds, 
needs thorough theoretical studies to clarify  whether
such cold ultrarelativistic winds can be formed and survived 
around supermassive black holes in the cores of AGN.

In this paper we suggest a new scenario which allows  
formation of  very (in practice, arbitrary) hard gamma-ray spectra
in a quite natural way. The model is based on a 
postulation that gamma-rays before leaving the source 
suffer significant photon-photon absorption due to interactions  
with dense radiation fields inside or in the vicinity of 
compact gamma-ray production region(s).  
Interestingly,  the presence of high density 
radiation fields of different origin in the inner parts 
of blazars generally is treated as a problem for 
the escape of high energy gamma-radiation from their production region,
and, in this regard, the current models of TeV blazars are designed 
in a way to avoid the internal gamma-ray absorption. Below we show
that, in fact,  a moderate internal photon-photon 
absorption can be a clue to the  very hard 
intrinsic energy spectra of TeV blazars. 
 
\section{Internal absorption of gamma-rays in blazars}

When propagating through an isotropic source of low-frequency 
radiation, the gamma-ray absorption at photon-photon 
interactions is characterized by the optical depth
\begin{equation}
\tau(E)=\int_0^R \int_{\epsilon_1}^{\epsilon_2} 
\sigma(E, \varepsilon) 
n_{\rm ph}(\varepsilon, r) {\rm d} \varepsilon {\rm d} r \ ,
\label{optdepth}
\end{equation}
where $n_{\rm ph}(\varepsilon, r)$ describes the spectral and 
spatial distributions of target photons in the source
of size $R$.  
With a good accuracy,
the total cross-section in the monoenergetic isotropic   
photon field can be represented in the form
(see e.g. Aharonian 2004):
\begin{eqnarray}
\sigma_{\gamma \gamma}  & =&
\frac{3 \sigma_{\rm T}}{2 s^2}
\left[ \left(s+ \frac{1}{2} \ln s- \frac{1}{6} +\frac{1}{2s} \right)
\ln(\sqrt{s}+ \sqrt{s-1})  -  \right. 
\nonumber \\
 & & \left.  
\left(s+ \frac{4}{9} - \frac{1}{9s}\right)  \sqrt{1- \frac{1}{s}}\right] \ . 
\label{crosssection}
\end{eqnarray}
The cross section  depends  only on the product of the 
primary ($E$) and target photon ($\varepsilon$) energies,  
$s=E \varepsilon/m_e^2 c^4$. 
Close to the threshold,  $s  \rightarrow  1$, 
the pair production cross-section behaves as   
$\sigma_{\gamma \gamma} \approx (1/2) \sigma_{\rm T} (s-1)^{3/2}$.
The cross-section decreases  with $s$ also when $s  \gg 1$:
$\sigma_{\rm \gamma \gamma} \approx (2/3) \sigma_{\rm T} s^{-1} \ln s$. 
The cross-section achieves its maximum at $s \approx 3.5$:
$\sigma_{\gamma \gamma}  \approx 0.2  \sigma_{\rm T}$. 

For a homogeneous source  with a narrow spectral distribution
of photons,   for order of magnitude estimates one can use 
the   approximation 
$\tau(E) \simeq R \sigma_{\gamma \gamma}(E, \bar{\varepsilon}) 
n(\bar{\varepsilon})$. In this case we should expect maximum absorption effect 
at gamma-ray energy $E^\star \approx m_{\rm e}^2 c^4/\bar{\varepsilon}$. 
Both at lower and higher energies, the source becomes more transparent, thus 
we should expect a quite strong deformation of the primary spectrum. 
In the case  $\tau(E^\star) \geq 1$,
the effect could be dramatic, given the exponential dependence of 
the absorption on the optical depth. Note that while 
for a narrow spectral distribution of target photons
the monoenergetic approximation gives a quite accurate estimate 
of the effect at $E \gg E^\star$, 
at low energies, $E \leq 1/4 E^\star$, this  approximation  
implies a completely transparent source (i.e. $\tau=0$) 
although, a non-negligible  absorption can take place also 
below $E^\star$.  
For example,  because of interactions with the Wien tail,
the absorption effect in the black-body radiation field 
cannot be disregarded even at very low energies,  
$E \ll  m_{\rm e}^2 c^4/kT$. 

\begin{figure}
\includegraphics[width=0.4\textwidth]{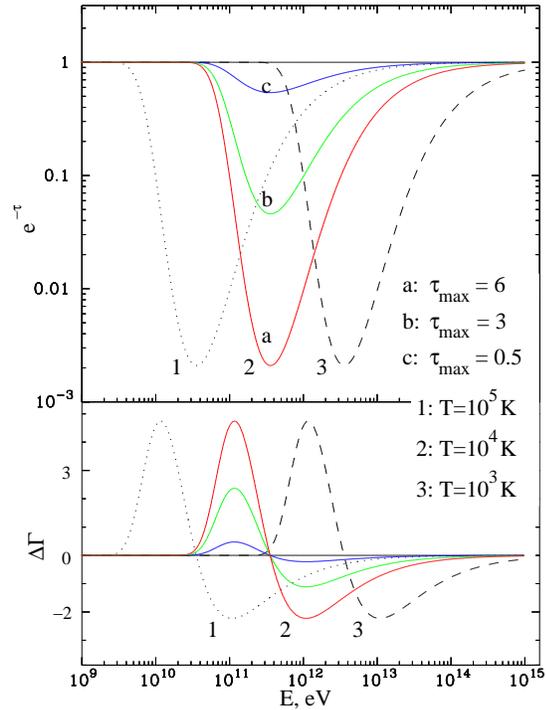}
\caption{\textit{Upper pannel}: Attenuation factor 
$\kappa=\exp{(-\tau)}$ for three different temperatures of target photons, 
$T=10^3$~K, $10^4$~K, and $10^5$~K (for the 
optical depth $\tau_{\rm max}=6$). For $T=10^4$~K calculations are performed for three different optical depths: 
$\tau_{\rm max}=$ 0.5 (blue), 3 (green), 6 (red). \textit{Bottom pannel}: 
Variation of the local photon index
of the gamma-ray spectrum.}
\label{atten:bb}
\end{figure}

In Fig.\ref{atten:bb} (upper panel) we present the gamma-ray attenuation factor,
$\kappa=\exp{(-\tau)}$
in a grey-body radiation field described by Planckian distribution with 
three different temperatures $T=10^3$~K, $10^4$~K, and $10^5$~K 
calculated for an optical depth fixed
at the energy corresponding to the maximum absorption, 
$E^\star \approx m_e^2 c^4/kT$ \cite{Gould}: 
Since the optical depth is a function of 
the product $E \cdot kT$, three curves are identical, but 
shifted relative to each other by a factor proportional 
to the radiation temperature.  
In Fig.\ref{atten:bb}   we show also the attenuation 
factors for a fixed temperature $T=10^4 \ \rm K$ calculated for 
three different  optical depths  
$\tau_{\rm max}=$ 0.5, 3, and 6. It is 
seen that the  gamma-ray attenuation, starting 
from the energy $E \sim 0.1 E^\star$, gradually increases
up to $E \sim E^\star$, after which   
the source becomes more and more transparent 
($\kappa \propto \exp{[-E^{-1} \ln E]} \rightarrow 1$),
and, consequently,  the primary spectrum starts to recover. 
As a result, in this energy interval the spectrum appears
harder than the primary spectrum. The bottom panel of 
Fig.\ref{atten:bb} shows the change of the slope
of gamma-ray spectrum, $\Delta \Gamma$,
which can be interpreted as a change of the local photon index,
assuming that the initial gamma-ray spectrum
is  described by a power-law distribution, 
${\rm d}N/{\rm d}E \propto E^{-\Gamma_0}$. 
The emerging spectrum of gamma-rays  in the energy interval 
$(0.1-1) E^\star$ is steeper than the 
initial spectrum ($\Delta \Gamma \ge 0$); 
it recovers at $E=E^\star$ ($\Delta \Gamma=0$), and at energies  
$E \ge E^\star$ the spectrum becomes significantly harder than the 
initial spectrum ($\Delta \Gamma \le 0$). 
For example, in the case of 
initial gamma-ray spectrum $E^{-2}$ and $\tau_{\rm max}=6$, 
the emerging spectrum of gamma-rays  in the energy interval 
$(1-10)E^\star$ can be  very hard with $\Gamma=\Gamma_0 + \Delta \Gamma \sim 0$, 
although the absolute flux is suppressed by an almost three orders of magnitude  at 
$E=E^\star$, and an order of magnitude at  $E=10E^\star$. 

\begin{figure}
\includegraphics[width=0.4\textwidth]{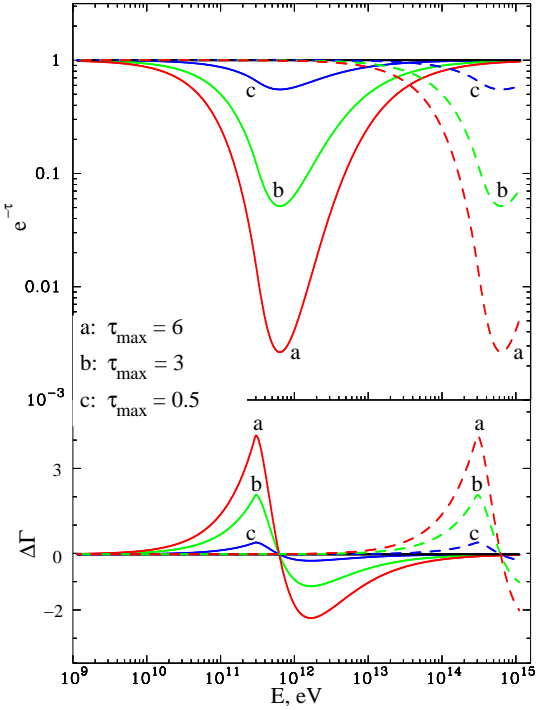}
\caption{\textit{Upper pannel}: Attenuation factor calculated 
for a power-law distribution of target photons with a low-energy cutoff:
$n_{\rm ph}\propto\varepsilon^{-2}$ for $\varepsilon_1<\varepsilon<\infty$
and $n_{\rm ph}=0$ for $\varepsilon \le \varepsilon_1$. 
Solid curves: $\varepsilon_1=1$~eV; dashed curves: 
$\varepsilon_1=10^{-3}$~eV. Three different optical depth are shown: $\tau_{\rm max}=$ 0.5~(blue), 3~(green), 6~(red). \textit{Bottom pannel:} Variations of the 
local photon index of the gamma-ray spectrum.
\label{atten:pp}}
\end{figure}

Note that the requirement of a narrow  spectral distribution of 
the target photons is a key condition  
for this remarkable effect. It should not necessarily be 
a Planckian or monoenergetic distribution, but may have any other shape,
for example power-law with a low-energy cutoff: 
$n(\varepsilon) \propto \varepsilon^{-\alpha}$ at 
$\varepsilon \geq \varepsilon_1$, and $n(\varepsilon) =0$ at 
$\varepsilon \leq \varepsilon_1$. In this case, the low-energy cutoff  
$\varepsilon_1$ plays a similar role as the  
temperature in the Planckian distribution. 
This is demonstrated in Fig.~\ref{atten:pp} 
for a power-law distribution of the 
background field with $\alpha=2$ and sharp cutoff 
at  $\varepsilon_1=1 \ \rm eV$ and $10^{-3} \ \rm eV$.  
Indeed, for the same $\tau_{\rm max}=6$, 
the case of $\varepsilon_1=1 \ \rm eV$ 
is quite similar  to the case of Planckian distribution with 
temperature $T=10^4$~K  shown in Fig.\ref{atten:bb}. 
The main difference appears in the low-energy part, 
$E \ll E^\star$. The absorption curve in 
Fig.\ref{atten:pp}  at low energies is smoother,  because 
the $n(\varepsilon) \propto \varepsilon^{-2}$ type distribution provides 
more high-energy target photons compared to the Wien tail of
the thermal distribution, for interactions with low-energy gamma-rays.

\begin{figure}
\includegraphics[width=0.4\textwidth]{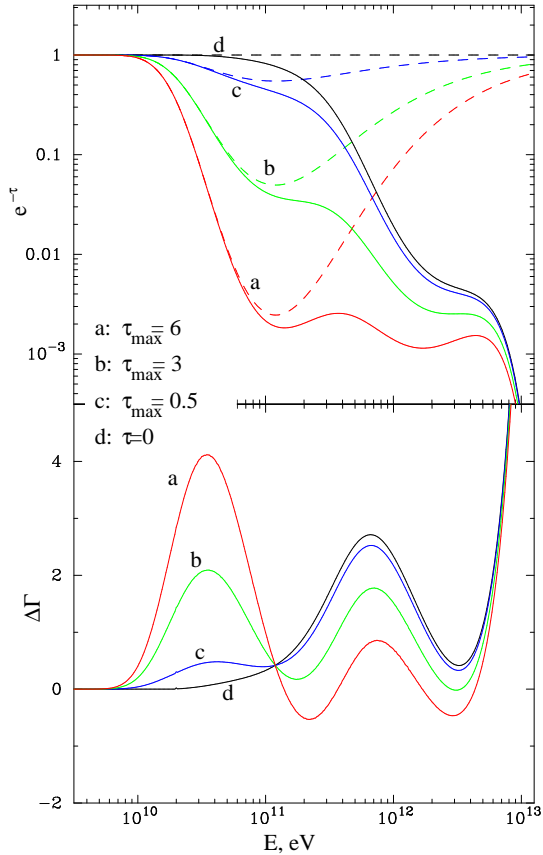}
\caption{Internal and intergalactic absorption of gamma-rays. 
\textit{ Upper pannel:} Attenuation factors.
The internal absorption is calculated for a Planckian distribution
of target radiation with temperature $T=5 \times 10^4$~K. Three dashed 
curves correspond to the internal optical depths $\tau_{\rm max}=$ 6~(a) 
3~(b), 0.5~(c), 0~(d).  
The corresponding solid curves include both the 
internal and intergalactic absorption. The intergalactic 
absorption is calculated for a source at $z=0.186$ (the redshift
of the BL Lac object 1ES~101-232), assuming a reference shape of the EBL spectrum 
close to the one calculated by Primack et al. (2005), and normalized to the EBL flux at
$2.2 \ \mu \rm m$:
$u_{\rm EBL}(2.2 \ \mu \rm m)=16 \ \rm nW/m^2 sr$.  
\textit{Bottom pannel:} Variation  of the local photon index.}
\label{2abs:a}
\end{figure}

\begin{figure}
\includegraphics[width=0.4\textwidth]{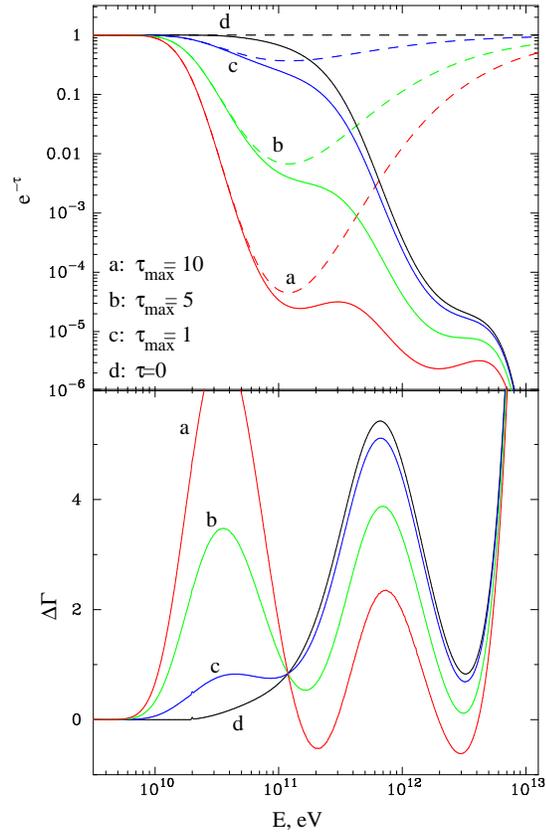}
\caption{The same as in Fig.\ref{2abs:a}, but the  curves are calculated 
for the internal optical depths 10~(a), 5~(b), 1~(c) and 0~(d). 
For EBL flux is assumed twice larger than in Fig.\ref{2abs:a}: 
$u_{\rm EBL}(2.2 \ \mu \rm m)=32 \ \rm nW/m^2 sr$.
\label{2abs:b}}
\end{figure}

The hardening of the initial spectrum caused by internal absorption 
compensates, to a large extent, the steepening of the spectrum 
due to intergalactic absorption. This is demonstrated in 
Figs.\ref{2abs:a} and \ref{2abs:b}. 
For the EBL flux we us a ``reference'' shape close to the one 
calculated by Primack et al. (2005), but  with two different absolute 
flux normalizations at the wavelength $\lambda=2.2 \mu \rm m$: 
$u_{\rm EBL}(2.2 \ \mu \rm m)=16 \ \rm nW/m^2 sr$ (Fig.\ref{2abs:a}) and 
$32 \ \rm nW/m^2 sr$ (Fig.\ref{2abs:b}). The first flux is a factor of two larger than the low-limit of EBL corresponding to the integrated 
light contributed by resolved galaxies
\cite{Madau}, while the second flux can be treated as an 
upper limit at $2.2 \mu \rm m$ (it is 
slightly higher than 
the fluxes claimed  from the  
COBE/DIRBE and 2MASS 
measurements \cite{Wright,Cambresy}.  Note that for the 
primary (unabsorbed) differential gamma-ray spectrum 
with a photon index $\Gamma_0=2$, the attenuation 
factor $\kappa=\exp{(-\tau)}$ describes the 
SED  of the absorbed 
radiation ($E^2 {\rm d}N/{\rm d}E \propto \kappa(E)$). 
All curves are obtained for a source located at $z=0.186$. 
This is the redshift of the BL Lac object 1ES~101-232, the gamma-ray 
observations of which by the HESS collaboration  
have been initially used to constrain the EBL flux 
at optical and NIR wavelengths \cite{1101_HESS}. 

The solid curves 
 in  Figs.\ref{2abs:a} and \ref{2abs:b} (marked as ``d'') 
correspond to the pure intergalactic effect (i.e. without 
the internal absorption). They show 
strong steepening of the spectrum below 1 TeV and above 10 TeV with a noticeable recovery of the initial shape around a few TeV, which is explained by the specific shape of the EBL energy flux  
(close to $u_{\rm EBL} \propto \lambda^{-1}$)
between 2 and 10 $\rm \mu m$ \cite{Hamburg01}. It is seen that for the 
absolute flux of EBL with a normalization at 
$2.2 \mu \rm m$, $u_{\rm EBL}(2.2 \ \mu \rm m)=16 \ \rm nW/m^2 sr$,
the photon index between
100 GeV  and  several TeV is changes by 
$\Delta \Gamma \sim 2$, thus, 
the intrinsic (source) spectrum should be very hard with a 
slope $\Gamma = \Gamma_{\rm obs} - \Delta \Gamma \sim 1$,
where $\Gamma_{\rm obs}=2.88 \pm 0.17$ is the observed photons 
index of 1ES~101-232 \cite{1101_HESS}. Postulating that the photon index
allowed by  conventional models of gamma-ray production in blazars 
should not exceed  $\Gamma_0 = 1.5$, an upper limit on the EBL flux, 
at the level of  
$u_{\rm EBL}(2.2 \ \mu \rm m) \approx 10 \ \rm nW/m^2 sr$
has been derived by the HESS collaboration \cite{1101_HESS}. 
It is seen from Fig.\ref{2abs:a} that this upper limit 
can be readily increased by a factor of 1.5, 
alowing a substantial internal absorption of gamma-rays. 
Indeed,  the ``joint operation'' of internal and intergalactic absorptions
results in the changes of the slope of the initial gamma-ray spectrum  
in the relevant energy band by 
$\Delta \Gamma \sim 0$ to 1.5 for $\tau_{\rm max}=3$, 
and $\Delta \Gamma$  from  -0.5 to  +0.5  for $\tau_{\rm max}=6$.  
In the latter case, the overall change of the shape of the initial spectrum from 100 GeV to 5 TeV is quite small, so hard TeV gamma-ray 
spectra   can in principle  be detected also from  distant blazars. 
This assumption can solve, to a certain extent,
the problem related to the spectra of accelerated (parent) particles,
and thus (unfortunately!) relax  the constraints on the 
EBL that can be derived from gamma-ray observations.
Formally, a detection of not-very-steep TeV gamma-ray spectra with $\Gamma_{\rm obs} \leq 4$  
from distant ($z \geq  0.15$) blazars
cannot be excluded even  for an EBL flux close to the 
claimed high EBL fluxes derived from the   COBE/DIRBE and 2MASS
measurements \cite{Cambresy}, $u_{\rm EBL}
(2.2 \rm \mu m) \approx 28 \ \rm nW/m^2 sr$. 
This is demonstrated in Fig.\ref{2abs:b}.  At  energies above 300 GeV 
the absorption of gamma-rays in EBL only increases the photon
index  by $\Delta \Gamma \geq 4$,  while an  additional
internal absorption with $\tau_{\rm max}=10$  results in 
$\Delta \Gamma \leq 2$.  Approximately the same result is expected 
for low EBL (e.g. at the level of 
$u_{\rm EBL}(2.2 \rm \mu m)=10 \ \rm nW/m^2 sr$), but
for sources located at $z \gg 0.1$. In this regard, the recent 
claim of detection 
of VHE gamma-rays from 3C~279 ($z=0.538$)  
by the MAGIC collaboration \cite{3C279}
can be an indication of significant gamma-ray 
absorption inside the source. 
Otherwise,  even for the minimum possible 
EBL flux,  the pure intergalactic absorption 
would result in an extremely steep 
VHE gamma-ray spectrum with a photon index $\geq 5$. 
It is intersting to note that the UV photons of the 
broad-line emission region of a size 
$R \sim 10^{17} \ \rm cm$ and 
luminosity $L \sim 2 \times 10^{44} \ \rm erg/s$ \cite{Pian}
can serve as a perfect target for internal
absorption of high energy gamma-rays in 3C~279.  

%

The internal absorption of gamma-rays significantly  
increases the  energy requirements to the source. For example 
in the case of $\tau_{\rm max}=5$, the internal absorption 
leads to the reduction of the observed flux at 1 TeV by an
additional factor of 10. This, combined with 
the intergalactic absorption,
implies 3 orders of magnitude attenuation of 
the primary radiation (see Fig.\ref{2abs:a}).  
For  the quiescent state of 1ES~101-232, 
the corresponding apparent  gamma-ray luminosity around 1 TeV 
is estimated $L_{\rm TeV} \simeq  10^{44} \kappa^{-1} \ \rm erg/s$.
For the attenuation factor $\kappa(1 \ \rm TeV) \sim 10^{-3}$,
it becomes huge, $10^{47}  \ \rm erg/s$, and perhaps one or two orders
of magnitude even larger in the flaring states (in analogy with 
Mkn 421, Mkn 501 and PKS 2155-304) \footnote{The detected 
gamma-ray flux  around 200 GeV is an order of magnitude 
larger than at  1 TeV, however, because of 
dramatic reduction of the absorption effect, the contribution 
of low energies to the (absorption-corrected)  apparent luminosity, 
is relatively small.}. Nevertheless, since there is little doubt that 
gamma-rays are produced in relativistically moving jets 
with a Doppler factor $\delta \geq 30$ or even $\geq 100$, as it 
follows from the recently reported  variability of Mkn 501 \cite{Mkn501_Magic} and 
PKS 2155-304 \cite{PKS_HESS} on minute scales  (see e.g. Fabian et al. (2007)),
the intrinsic luminosity $L_{\rm int}=L_{\rm app} \delta^{-4}$,  
could be quite modest, namely, 
at the level of $10^{39} - 10^{41} \ \rm erg/s$ 
which, in fact, is not far from the TeV gamma-ray 
luminosity of the nearby non-blazar type AGN  M87 \cite{M87_HESS}.

The attenuation  of TeV gamma-rays  dramatically increases  
the intrinsic TeV to GeV gamma-ray flux ratio. This  
does not, however, contradict the GeV 
flux upper limits available from the EGRET observations 
(typically, at the level of $10^{-10} \ \rm erg/cm^2 s$ or higher), 
especially if one takes into account that  
the energy spectra of gamma-rays produced 
in some principal radiation processes (e.g. through inverse 
Compton scattering or proton synchrotron radiation) 
at sub-TeV energies are  expected harder than $E^{-2}$. On the other 
hand, we certainly expect GeV gamma-rays 
from TeV blazars, and in this respect, 
the upcoming GLAST measurements with significantly 
improved (compared to EGRET) sensitivity,
especially at multi-GeV energies, should provide the first 
effective probes of TeV blazars in the MeV/GeV domain in general,
and for the ``internal gamma-ray absorption'' scenario, in particular.   

Finally, in the case of hadronic origin of TeV 
gamma-rays produced at 
$pp$ and/or $p\gamma$ interactions, the flux of accompanying 
TeV neutrinos, which  freely penetrate through the 
internal and extragalactic radiation fields,  
can be as high as $10^{-10} \ \rm neutrinos/cm^2 s$,
i.e. above the detection threshold of the   
next generation km$^3$ scale neutrino detectors. 
The detection of both gamma-rays and neutrinos from TeV blazars, 
and the comparison of fluxes of these two components of radiation
would provide principal information about the high energy processes 
in blazars, as well as about the attenuation of gamma-rays due 
to the (combined) internal and intergalactic absorption. 
An additional information about 
the internal photon-photon absorption alone 
(separated from the extragalactic absorption) 
is contained in the radiation of secondary (pair-produced) electrons.

\section{Radiation of secondary electrons}

The propagation of high energy gamma-rays through 
a  low-energy photon field cannot be 
reduced to the simple  effect of absorption. When 
the gamma-ray photon is absorbed,  its energy  
is transfered to the electron-positron  pair. The secondary electrons 
interacting with the ambient magnetic and radiation fields produce high
energy photons, either via synchrotron radiation or inverse Compton scattering. 
The synchotron photons are produced with much smaller energies,  
thus they do not interact with the 
background  low-energy  photons. In a target 
field with narrow spectral distribution, 
inverse Compton scattering of the photo-produced 
electrons proceeds in the Klein-Nishina limit; 
the upscattered photon receives the major fraction of the electron energy, 
thus is  able to interact 
again with background photons. The second generation pairs again 
produce gamma-rays, thus an electromagnetic cascade developes.

%
\begin{figure}
\includegraphics[width=0.4\textwidth,angle=0]{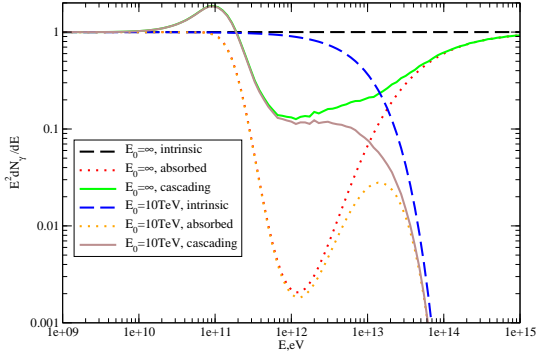}
\caption{Gamma-ray spectra caused by the internal 
photon-photon absorption (dotted curves) 
or formed  during a development of pair cascades (solid curves)
in a radiation field with temperature  
$T=10^4$ and  optical depth  $\tau_{\rm max}=6$.  
It is assumed that the energy density of radiation significantly  exceeds
the energy density of the magnetic field, $u_{\rm r} \gg u_{\rm B}$.  
The  spectra  of primary gamma-radiation (dashed lines) 
are assumed as ``power-law with exponential cutoff'': 
${\rm d}N/{\rm d}E \propto E^{-2} \exp{(-E/E_0)}$ 
with   $E_{\rm 0}=10$~TeV and  $E_{\rm 0}=\infty$.
\label{cascade}}
\end{figure}
%

While the energy of gamma-rays interacting with EBL dissipates in 
the intergalactic medium, and in this way contributes to the diffuse 
extragalactic background radiation, the secondary radiation caused by internal 
absorption may accompany the primary (unabsorbed)  fraction 
of gamma-rays. In this regard, the development  of  an
electromagnetic cascade  is not a desirable process, 
because it masks the distinct  absorption features,
and thus prevents the formation of very hard gamma-ray spectra. This is 
demonstrated in Fig.\ref{cascade}. 
The cascade spectra are calculated 
assuming that the region of production
of primary gamma-rays is located in the center of a spherical source filled with 
grey-body radiation with temperature $T=10^4 \ \rm K$ and optical depth 
$\tau_{\rm max}=6$. The spectrum of primary gamma-rays is given  
in the  form: ${\rm d}N/dE \propto E^{-2} e^{-E/E_0}$ 
for two values of the cutoff energy: $E_{\rm 0}=10$~TeV and  $E_{\rm 0}=\infty$. 
In Fig.~\ref{cascade} both the absorbed and 
cascade gamma-ray spectra are shown. It is seen that the development 
of pair cascades fully washes out the absorption features, and instead 
forms  a standard spectrum with a maximum (``bump'')
around the 
interaction threshold, $\sim (m_e c^2)^2/kT \sim 100 \ \rm GeV$,  followed by  
a steep spectrum  above the ``bump''. The cascade spectrum around 
1~TeV saturates at the  level of 10 percent  of the primary gamma-ray flux. 
The spectrum above 1 TeV  has a flat shape until the efficiency of the 
cascade drops (because of the progressively 
decreasing optical depth) with a gradual transition 
to the ``absorption'' regime. 

There are two ways of significant reduction of the contribution from 
the cascade component to the observed gamma-radiation.  

\vspace{2mm}
\noindent
(i) \textit{Absorption of gamma-rays outside the production region}. 
In this case the ``foreground'' cascade 
radiation cannot screen the unabsorbed fraction of gamma-rays. Indeed, although 
the energy in the cascade radiation exceeds, by a factor of 100,
the energy of the unabsorbed fraction of primary 
radiation (see Fig.~\ref{cascade}), for the  observer  
the primary radiation 
emitted by a relativistically moving source ('blob``) 
will be much brighter 
($\propto \delta_j^4$) and shifted to higher energies ($\propto \delta_j$)
compared to the isotropic cascade emission of the surrounding environment. 
On the other hand, the part of the cascade developed inside the blob
is, of course, also Doppler boosted, therefore the condition of suppression of the 
cascade component can be satisfied when the optical depth 
within the blob $\tau^\prime \ll 1$, i.e. the 
gamma-ray production region is much smaller than 
the source of the optical/UV radiation, 
$l \ll  R$. Since th gamma-ray production 
regions in  blazars  are  believed to be very compact, 
$l \sim  10^{14} - 10^{16} \ \rm cm$,  we may conclude 
that the source of the optical radiation should be larger than $R \sim 10^{15}-10^{17}$ cm.
There are many potential sources of optical radiation in blazars (see e.g. 
Urry and Padovani 1995). In this  paper we do not intend to specify the origin of 
the low-energy radiation fields, but simply notice that an even
modest optical source located in the core of a blazar can provide an effective target 
for gamma-ray absorption. The luminosity of this source is estimated as 
$L_{\rm O}=12 \pi n_{\rm ph} kT R^2 c$, where $n_{\rm ph}$ 
is the average number density of radiation. It can be estimated from the condition 
$\sigma_{\gamma \gamma} R n_{\rm ph}=\tau_{\rm max}$. For a characteristic 
optical depth  $\tau_{\rm max} \sim 5$ and  temperature $T=5 \times 10^4$~K, 
we obtain, $L_{\rm O} \simeq 2 \times 10^{43} (R/10^{17} \ \rm cm) \ \rm erg/s$. 

%
\begin{figure}
\includegraphics[width=0.4\textwidth,angle=0]{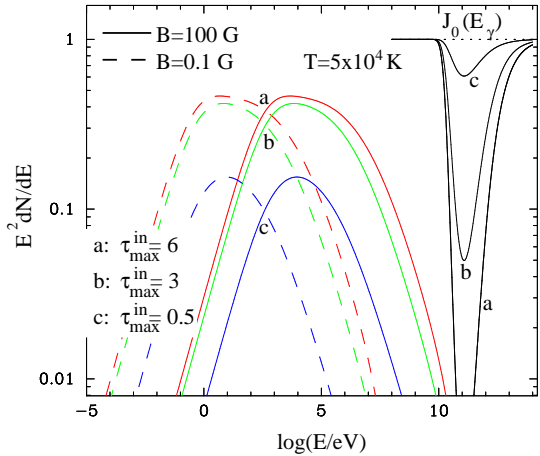}
\caption{Synchrotron radiation of secondary electrons. The primary gamma-ray spectrum is
assumed ${\rm d}N/{\rm d} E  \propto E^{-2}$ (dotted line). The target photon field is Planckian with $T=5\cdot10^4$~K. The absorbed gamma-ray spectra and the corresponding 
spectra of secondary electrons are calculated 
for three optical depths $\tau_{\rm max}^{\rm in}=$ 6~(a), 3~(b), and 0.5~(c).  
The  synchrotron radiation is calculated assuming that the absorption of gamma-rays 
takes place inside the source (gamma-ray production region) for two 
values of the magnetic field:  $B=100$~G (solid curves) and $B=0.1$~G (dashed curves).
\label{synch}}
\end{figure}

\vspace{2mm}
\noindent
(ii) \textit{Secondary electrons cooled through synchrotron radiation.} 
This condition can be satisfied if the energy density of the magnetic field exceeds the energy density of radiation,  
$B^2/8 \pi \geq 3kT n_{\rm ph}$ or 
$B \geq 0.4 (R/10^{17} \ \rm cm)^{-1/2} \ \rm G$.  
In Fig.~\ref{synch} the broad-band SED of the 
radiation initiated by absorption of primary gamma-rays with a power-law  
spectrum ${\rm d}N/{\rm d} E \propto E^{-2}$ is shown, 
assuming that the absorption of gamma-rays takes place inside the 
gamma-ray production region. Calculations are performed for 
three different optical depths $\tau_{\rm max}$=0.5, 3, and 6, 
a fixed temperature of radiation  $T=5\cdot10^4$~K, and two 
extreme values of the magnetic field,  $B=100$~G and $B=0.1$~G. 
In calculations of the electron spectra we assume that 
the energy losses of electrons are dominated by  
synchrotron cooling. Note that the spectrum of 
synchrotron radiation of secondary electrons 
has a characteristic bell-type form, i.e.  
quite similar to the synchrotron spectra  formed in the 
synchrotron-self Compton (SSC) models. However, 
in the SSC models designed  for TeV blazars 
the synchrotron peak is a result of the 
maximum energy of accelerated electrons,
while the hard low-frequency part of the spectrum is determined by 
radiation of uncooled low-energy electrons. 
In the ''internal gamma-ray absorption`` 
scenario we see similar features, but for 
different reasons. The electrons in a narrow-band radiation field 
are produced with a spectrum similar to the spectrum of parent gamma-rays 
($\propto E^{-\Gamma_0}$),
but with cutoffs both at low- and high energies. These cutoffs are 
explained by the threshold of the photon-photon pair production 
and by the reduction  of its cross-section at highest energies, 
respectively. Due to the 
synchrotron cooling, at low energies the electron spectrum obtains 
a standard $\propto E^{-2}$ form,  which gradually transforms to a 
$\propto E^{-(\Gamma_0+1)}$ type spectrum at intermediate energies and a 
cutoff  at highest energies. The corresponding spectral features 
are reflected in the synchrotron spectrum - a power-law 
with a photon index 1.5 at low energies, with a smooth transition to a 
power-law with a photon index $\Gamma_0/2 +1$ at 
intermediate energies, and a smooth
gradual  cutoff  at highest energies. 
A significant difference between the SSC and the ''internal gamma-ray absorption``
models  appears also in the ratio of fluxes corresponding to the 
low (IR to X-ray) and high (gamma) energy peaks. 
While in the SSC scenario this ratio is determined by the 
ratio of energy densities of the magnetic and target photon fields, 
in the ''internal gamma-ray absorption`` scenario this ratio is basically determined 
by the efficiency of gamma-ray absorption inside the source. The magnetic field
determines only the position of the synchrotron peak, but not the flux 
level. The latter  is determined  by the factor proportional to 
$(1 - \exp{[-\tau_{\rm max}]})$. For  $\tau_{\rm max} \gg 1$, the dependence 
on the optical depth almost disappears. 
All these features can be seen in Fig.~\ref{synch}. 

\begin{figure}
\includegraphics[width=0.4\textwidth]{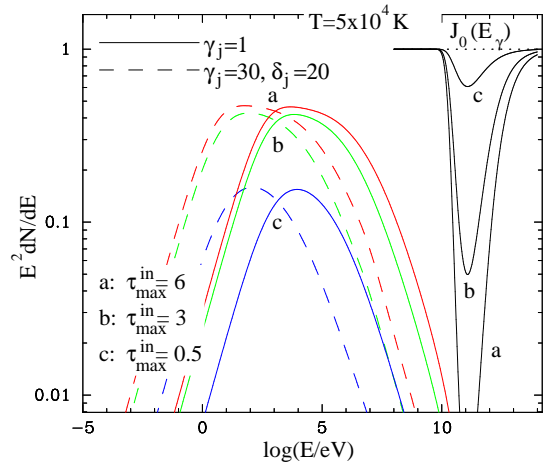}
\caption{Impact of the bulk motion on synchrotron radiation of secondary electrons.  The magnetized blob moves with a relativistic velocity.
The synchrotron maximum moves  by a factor of $\delta_j/\gamma_j^2$ and the distributions become somewhat wider. The primary gamma-ray spectrum was assumed power-law: $\propto E_\gamma^{-2}$ (is 
shown by the dotted line). The target photon field is Planckian with $T=5\cdot10^4$~K. The absorbed spectra are shown by black solid curves for $\tau_{\rm max}=6$(a), $\tau_{\rm max}=3$(b) and $\tau_{\rm max}=0.5$(c). We calculate synchrotron radiation from a magnetized region with the average optical depth (over different directions) $\tau_{\rm max}^{\rm in}=$ 6(a, red lines), 3(b, green lines) and 0.5(c, blue lines). The magnetic 
field is assumed $B=100$~G. Calculations are  performed for two different bulk Lorenz factors: $\gamma_j=1$(no motion) and $\gamma_j=30$ (for $\delta_j=20$).\label{Doppler}}
\end{figure}

The position of the synchrotron peak depends, in a quite interesting way, also on the 
Lorenz and Doppler factors of the source. While in the standard models of blazars 
the Doppler effect shifts 
the overall SED towards higher energies by a factor of $\delta_j$, in the 
''gamma-ray absorption``
scenario the synchrotron peak is shifted 
towards lower energies (see Fig.\ref{Doppler}) 
The reason is quite simple. In the frame of the relativistically moving 
source illuminated by an external radiation with a characteristic energy 
$\varepsilon_0$, the electrons are produced with energies 
$E_{\rm min}  \geq  (m_e  c^2)^2/(\varepsilon_0 \gamma_j)$, thus  
the position of the synchrotron peck in the frame of the observer is 
proportional to  $\delta_j E_{\rm min}^2 \propto \delta/\gamma_j^2$. Thus in a source 
moving towards the observer at a small angle, the position of the 
synchrotron peak in the frame of the observer 
is inversely proportional to the Doppler factor $\delta_j$, just opposite to
the spectrum of gamma-rays which is shifted towards higher energies by the same 
Doppler factor.

%

\begin{figure*}
\includegraphics[width=0.8\textwidth]{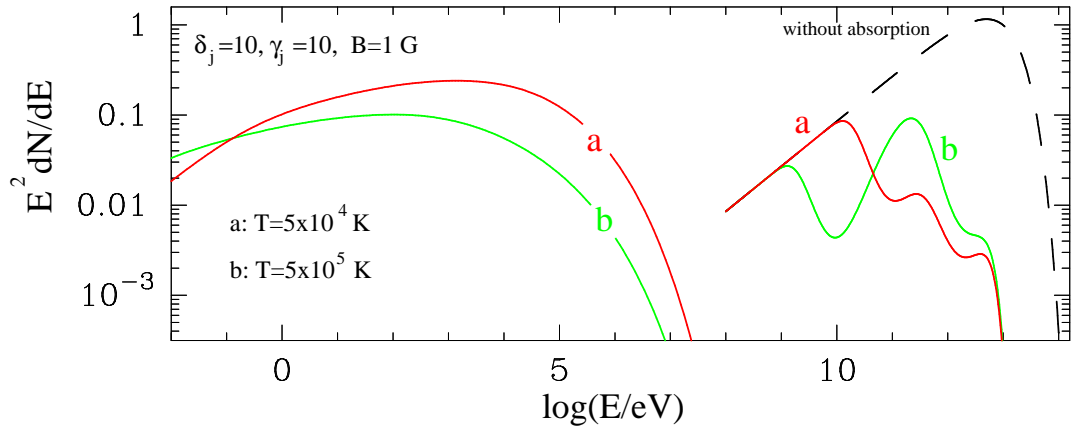}
\caption{Impact of the radiation temperature 
on the spectral energy distribution of emission 
initiated by primary gamma-rays in a jet moving through a homogeneous radiation field. 
The source is located at $z=0.186$. The dashed curve is the assumed 
primary spectrun of gamma-rays. The solid curves repreesent 
the synchrotron raduiation of the secondary electrons as well as 
the gamma-ray spectra after the internal and extragalactic absorption. 
Calculations are performed for two radiation temperatures,
$T=5 \times 10^4$~K (a) and $T=5 \times 10^5$~K (b). For both cases $\delta_j=\gamma_j=10$,
$B=1$~G, $\tau_{\rm max}=3$. An EBL is assumed with a normalization of the 
flux $u_{\rm EBL}(2.2 \mu \rm m)=16 \ \rm nW/m^2 str$. }
\label{temp}
\end{figure*}

The position of the synchrotron peak depends strongly 
also on the average energy (or temperature) of the 
target radiation field. Indeed, with an increase of the 
temperature of background radiation $T$, the threshold 
of photon-photon interactions, and consequently  the 
minimum energy of produced secondary electrons decreases  
as $E_{\rm min} \propto 1/T$, and hence the 
synchrotron peak moves towards lower energies 
as $h \nu_{\rm m} \propto 1/E_{\rm min}^2 \propto 1/T^2$. Generally,
the position of the synchrotron peak of pair produced electrons 
depends on the 
temperature of the target radiation field,  the magnetic field, 
and the Doppler and Lorentz factors of the jet,  as 
\begin{equation}
h \nu_{\rm m} \propto B T^{-2} (\delta_j / \gamma_j^2) \ . 
\end{equation}
It is easy to derive a simple  
analytical expression which 
described the high energy part of the synchrotron 
spectrum, $h \nu \gg h \nu_{\rm m}$, produced by 
electrons for which  the source becomes optically thin, 
$\tau(E) \le 1$. In this case the 
production spectrum of electrons 
$Q(E) \propto 1/E {\rm d}N_\gamma /{\rm d}E$ 
(here we ignore the weak logarithmic term in 
the photon-photon interaction cross-section). 
Then, for the power-law gamma-ray spectrum, 
${\rm d}N_\gamma /{\rm d}E \propto E^{-\Gamma_0}$, 
the cooled electron spectrum is also power-law, 
${\rm d}N_{\rm e} /{\rm d}E_{\rm e} 
\propto E^{-\Gamma_0-2}$, and correspondingly
the SED of synchrotron radiation, 
$\nu F_\nu \propto \nu^{-(\Gamma_0+1)/2}$. For example,
for $\Gamma_0=2$, the SED of synchrotron radiation 
is rather steep, $\nu F_\nu \propto \nu^{-1.5}$. In fact, 
because of the cutoff in the 
gamma-ray spectrum, the high energy tail of 
synchrotron radiation is expected even steeper. This is 
demonstrated in Fig.~\ref{temp} where the broad-band SEDs 
of radiation initiated by gamma-rays in a source at 
$z=0.186$ are shown. It is assumed that gamma-rays
in the frame of the jet moving with a 
Lorentz factor $\gamma_j=10$ have a power-law distribution with an
exponential cutoff at 1 TeV, 
${\rm d}N_\gamma /{\rm d}E \propto E^{-3/2} \exp{(-E/1 \rm TeV)}$.
It is assumed also that $\delta_j=\gamma_j$ (i.e. the 
viewing angle is $\theta \approx 6^\circ$).  The calculations 
are performed for two temperatures of the radiation field through 
which the jet propagates - $T=5 \times 10^4$~K and  
$T=5 \times 10^5$~K, assuming that in both cases  
the  optical depth inside the moving gamma-ray production region
(the blob) with a homogeneous magnetic field  
is $\tau_{\rm max}=3$. Finally for the intergalactic absorption 
a ''template`` EBL spectrum is assumed with a normalization at 
2.2 $\mu \rm m$ at the level of $16 \ \rm nW/m^2 ster$.  
The impact of the temperature 
on both the spectrum of arriving gamma-rays and the synchrotron 
radiation of secondary electrons is clearly seen. Note that 
while below 100 GeV the deformation of the primary 
gamma-ray spectrum is caused mainly by internal absorption, 
the sharp cutoff at energies above 10 TeV is due to the 
severe intergalactic absorption. In the intermediate energy 
range between 100 GeV and 10 TeV the internal and intergalactic 
photon-photon interactions ''operate`` together resulting in  
quite specific broad-band SEDs. 
The discussion of implications of these results for  
specific astrophysical objects is beyond the scope of this paper.
We note only that our preliminary studies show that the 
suggested model in general can satisfactorily explain the observed 
broad-band SEDs of TeV blazars.  

\section{Discussion} 

The energy spectra of VHE  gamma-rays from blazars, after correction  
for intergalactic absorption,  generally appear very hard, even  
for the minimum flux level of EBL determined by the integrated 
light of resolved galaxies at optical (Hubble)  and 
infrared (Spitzer) wavelengths.  A slight 
deviation from the  robust lower limits of EBL 
leads to unusually hard intrinsic gamma-ray spectra 
which cannot be easily explained within  the  
standard  particle acceleration and radiation models. 
In this paper we suggest a scenario which can lead to the 
formation of instrinsic gamma-ray spectra of arbitrary hardness
without introducing modifications in the 
particle acceleration models.
The main idea is that the gamma-rays  before they leave  the source
suffer significant internal energy-dependent absorption 
due to interactions with the ambient low-frequency photons. 
The existence of dense radiation fields of different origin in  
blazars (see e.g. Urry and Padovani (1995)) combined with the 
large photon-photon pair production cross-section, makes 
this scenario quite natural and effective, in particular in the 
compact cores of blazars. 
For the formation of hard VHE gamma-ray spectra,  
the target  radiation field must have a rather narrow 
spectral distribution or a sharp low-energy cutoff, 
with a typical energy of photons of about 1 to 10 eV.  Formally, for 
very large optical depths, this process can provide
an arbitrary hardness of gamma-ray spectra, though at the 
expense of a significant increase of the required 
nonthermal energy budget. 
However, as long as the current blazar models require 
relativistically-moving gamma-ray production 
regions with large Doppler factors,  
$\delta_j \geq 30$, and perhaps even more \cite{PKS_HESS,Fabian}, 
the available energy budget seems to be not a critical issue.

The unavoidable feature of the proposed model is the radiation of 
secondary electrons via synchrotron or inverse Compton scattering. 
If the optical depth inside the gamma-ray production region 
is small, $\tau \ll 1$, e.g. the gamma-ray source  is much smaller 
than the external source of optical photons, the secondary 
electrons are produced and radiate mainly outside the gamma-ray production region. 
Even in the case of heavy absorption of gamma-rays,
the secondary radiation of secondary electrons can hardly 
be detected.  
Indeed, since the intrinsic gamma-ray luminosity  
is relatively modest, and the absorbed energy is re-radiated 
as an isotropic source\footnote{Unless the electrons are produced in an environment 
with  a very low magnetic field, and thus are cooled via inverse Compton scattering 
before any noticeable deflection.}, the lost of the beaming factor dramatically 
reduces the  signal compared to the primary (Doppler boosted) radiation.  

The picure is dramatically changed when the gamma-ray source moves 
through a very dense photon field, such that the optical depth 
inside the source becomes larger than 1. In this case the main 
fraction of the absorbed energy is released in the form of secondary electrons
inside the gamma-ray production region, and thus the radiation 
of the secondary electrons profits,  
as the primary gamma-radiation does, from the Doppler boosting.  
The secondary electrons are cooled through synchrotron and/or inverse 
Compton channels. The latter in fact 
proceeds via development of pair cascades 
as long as the typical energies of electrons or gamma-rays  
and the energy of target photons $\varepsilon E_{\rm e, \gamma} \gg m^2_e c^4$. 
The cascade, however,  diminishes the energy-dependent absorption 
features, thus the  model becomes effective  when the electrons are
cooled predominantly via synchrotron radiation, i.e. 
$B^2/8 \pi \geq u_{\rm r}$. The energy density of the radiation 
$u_{\rm r} = \bar{\varepsilon} n_{\rm ph}$
with an average energy of target 
photons of about $\bar{\varepsilon} \sim 1 \ \rm eV$ is estimated 
from the condition $\tau_{\rm max} \geq 1$, thus for 
the effective suppression of the cascade 

\begin{equation}
 B \geq (40 \pi \bar{\varepsilon}/\sigma_{\rm T} l)^{1/2} 
\approx 0.5 (l/10^{15} \rm cm)^{-1/2} \tau^{1/2}_{\rm max}\ \rm G
\label{B}
\end{equation}
 
For an optical depth $\tau_{\rm max} \sim 1$ and 
the size of the gamma-ray source  $l \sim   10^{16} \ \rm cm$, the magnetic field 
exceeding  0.1 G should be sufficient to prevent the cascade. For smaller gamma-ray production regions, e.g. $l \sim 10^{14} \ \rm cm$, 
the magnetic field  should be  larger than 1 G.
For such magnetic fields the synchrotron radiation of secondary electrons 
appears in the  optical to hard X-ray energy bands.  Depending on the optical depth, 
the synchrotron peak can be higher than the gamma-ray peak.
Interestingly, unlike
the classical ''synchrotron/Inverse-Compton``  models, 
where the ratio of the synchrotron to IC 
peak is determined by the ratio $u_{\rm B}/u_{\rm r}$, 
in the ''internal gamma-ray absorption`` scenario
the synchrotron peak does not strongly depend on the magnetic field. 
Whether this scenario
can be applied to the the broad-band SEDs of gamma-ray blazars, 
is an interesting issue which requires special dedicated studies. 

Finally, we want to discuss briefly the radiation mechanisms of primary 
gamma-radiation. Generally, the model  does not give a preference to  
the leptonic or hadronic origin of radiation, unless the magnetic 
field exceeds the estimate given by Eq.(\ref{B}). In this case the synchrotron-to-IC flux ratio produced by directly accelerated 
electrons would be too high, especially after the internal absorption of gamma-rays, 
contrary to the detected SEDs of most of the TeV blazars. 

Large  magnetic fields in the gamma-ray production region, typically 
$B \geq 1 \ \rm G$, would favor gamma-ray production by  relativistic 
protons, with all advantages and disadvantages common for hadronic models. The 
basic problem of  hadronic models is linked to the low interaction 
rates which do not allow the most natural explanation of the observed 
fast gamma-ray variability of blazars in terms of 
radiative cooling. For example, in the case of interactions 
of protons with the ambient plasma with number density $n$, the characteristic 
time of $pp$ interactions with 
production of $\pi^0$-mesons is $t_{\rm pp} \approx 10^{15} n^{-1} \ \rm s$. Thus, in 
order to explain the variability of gamma-rays as short as several minutes
like the TeV flares observed from PKS 21555-301 and Mkn 501, 
the density of plasma should be 
as large as $5 \times 10^{12} \delta_j^{-1} \ \rm cm^{-3}$ which implies 
a very heavy source  and correspondingly  huge kinetic energy 
$E_{\rm kin}= (4/3) \pi l^3 n m_pc^2 \gamma_j  \approx 10^{55} \ \rm erg$ (here we assume that 
$\delta_ \approx \gamma_j$). One may invoke alternative explanations of 
the variability of blazars, e.g. due to the adiabatic losses or escape of 
particles from the source, but this assumption leads to dramatic 
reduction of radiation efficiency, and to an 
increase the energy requirements to the  accelerated protons. 

A similar problem face the photomeson processes at interactions of protons with the 
ambient radiation fields. Actually in the ''internal gamma-ray absorption`` scenario
this mechanisms seems a quite natural choice because  the same  background 
photons which absorb gamma-rays can play a role of the target for photomeson interactions. 
However, because of the small cross-section, the efficiency of this process 
again appears  quite low. 
The interaction time of protons with energy,   
$E \geq 200 \ \rm MeV/(\bar{\varepsilon} \gamma_j) \simeq  
2 \times 10^{16} (\bar{\varepsilon}/1 \ \rm eV)^{-1} (\gamma_j/10)^{-1} \ \rm eV$ 
(in the frame of the 
moving source  with a Lorentz factor $\gamma_j$)
is estimated  
$t_{\rm p \gamma} \approx 1/(f \sigma_{\rm p \gamma} n_{\rm ph} c) \sim 
(\sigma_{\rm \gamma \gamma}/\sigma_{\rm p \gamma})f^{-1}  R/c \tau_{\rm max}^{-1}$ 
($<\sigma_{\rm p \gamma}> \approx  10^{-28} \ \rm cm^2$ is the average cross-section and $f \sim 0.2$  is the multiplicity of the process). 
Thus, we can see that during the 
passage of the source of optical photons of size $R$, the protons transfer only  
$\sigma_{\rm p \gamma}/\sigma_{\gamma \gamma} \sim 10^{-3}$ fraction of their 
energy to gamma-rays. If such a low efficiency can be compensated by very  
large Doppler boosting (e.g. assuming $\delta_j \sim 100$), this channel can provide 
very large fluxes of neutrinos,
which unlike gamma-rays do not suffer internal and extragalactic absorption.
In the case of attenuation of VHE gamma-ray fluxes by 2 to 3 orders of magnitude, 
the expected fluxes of neutrinos from TeV blazars  can be as large as  
the detection threshold of the km$^3$ volume high energy neutrino telescopes,
$F_{\nu_\mu} (\geq 1 \ \rm TeV) \approx 10^{-11} \ \rm neutrinos/cm^2 s$. 
 
It is interesting to note that, because of the threshold of photomeson production, 
the interactions of protons of arbitrary 
distribution with a narrow band radiation with a characteristic energy  
$\bar{\varepsilon}$, result in a differential 
gamma-ray spectrum which below the energy 
$\approx 10^{16} (\bar{\varepsilon}/1 \ \rm TeV)^{-1}$ eV is extremely hard,
${\rm d}N/{\rm d}E=\rm const$, thus this process itself can provide very hard 
gamma-ray spectra independent of the spectrum of parent protons.  

Despite certain attractive features, this mechanism faces 
the same problem  as $pp$ interactions - a low 
radiation efficiency.
Therefore it can work only under conditions of extremely large 
Doppler boosting of radiation. The efficiency of VHE gamma-ray production  
can be much higher in the case of synchrotron radiation of protons, provided that
the acceleration of protons proceeds at a rate close to the fundamental limit, 
and  the magnetic field in the proton accelerator well exceeds 10 G. In particular,
in the magnetic field of order $100$ G, protons can be accelerated to energies 
$10^{20} \ \rm TeV$ and thus can produce VHE synchrotron gamma-rays on timescales of $10^{4}$ s. Although due to the  self-regulated synchrotron cutoff \cite{psynch}
the spectrum of gamma-rays is limited by sub-TeV energies, an  observer 
detects  Doppler boosted gamma-radiation extending to multi-TeV energies.  
The characteristic feature of this 
mechanism is the very large electromagnetic energy contained in the blob, 
$1/6 l^3 B^2 \approx  2 \times 10^{48} \ \rm erg$,  
hard X-ray emission of the secondary (pair-produced) electrons, and 
negligible fluxes of  neutrinos.


\hyphenation{Post-Script Sprin-ger}

\label{lastpage}

\end{document}